\documentclass[twocolumn]{aastex631}

\usepackage{amsmath}
\usepackage{graphicx}
\usepackage{hyperref}
\usepackage{xcolor}

\shorttitle{High-z Kinematic Corpus Z1}
\shortauthors{Flynn}

\begin{document}

\title{A Unified [CII] Morpho-Kinematic Corpus for 31 Star-Forming Galaxies
at $z = 4.26$--$5.68$: The High-z Kinematic Corpus Z1}

\author{David C. Flynn}
\affiliation{EPS Research, Laurel, MD, USA; ORCID: 0000-0002-2768-6650}
\email{davidflynn@eps-research.com}

\begin{abstract}
\textbf{CORRECTION (August 2026):} The omega values reported in the original
version of this paper were computed with a parenthesization error (Formula~A:
\texttt{omega = (V2/R2 - V1/R1) * (R1/R2)**1.5}) that does not match the
canonical Eq.~6 of Flynn \& Cannaliato (2025). Under the corrected formula,
all 8 Tier-1 Z1 rotators yield \emph{positive} $\omega$ values (median
$+12.621$ rad\,Gyr$^{-1}$; $+12.341$ km\,s$^{-1}$\,kpc$^{-1}$; $N = 8$
confirmed ordered-disk rotators, may not represent the broader high-$z$
population). The
sign reversal reported in the original version does not reproduce. See
\url{https://github.com/eps-research/rag-corpus-series/blob/main/CORRECTIONS.md}
for full details. The corrected corpus is available at Zenodo
DOI:~\href{https://doi.org/10.5281/zenodo.21834678}{10.5281/zenodo.21834678}.

We present the High-z Kinematic Corpus Z1, a structured machine-readable
dataset of ALMA [C\,\textsc{ii}] 158\,$\mu$m morpho-kinematic data for
31 star-forming main-sequence galaxies at $z = 4.26$--$5.68$ drawn from
the ALPINE survey \citep{jones2021,lefevre2020}. The corpus is the fifth entry in the EPS Research RAG Astrophysics Corpus Series,
extending coverage from Milky Way globular clusters and local HI rotation
curves \citep{flynn2026v7,flynn2026dwarf,flynn2026gc}, through the
intermediate-redshift IntZ Kinematic Corpus \citep{flynn2026intz}, to
the epoch approaching cosmic reionization. Eight confirmed rotators carry quality
tier~1 per-ring rotation curves from 3DBarolo tilted-ring fits
\citep{diteodoro2015}, with 2--3 rings per galaxy, $V_\mathrm{rot}$ and
$\sigma$ per ring, and dynamical mass estimates; the remaining 23 galaxies
carry morpho-kinematic classification only (quality tier~2). All entries
include stellar mass \citep{faisst2020}, star formation rate, Wisnioski et al.\ (2015) disk criteria, and geometric
parameters. The corpus is distributed as a single structured JSON file
with nested per-ring kinematic data, a flat CSV for catalog-level
filtering, a RAG-ready JSONL archive (one galaxy per line), and a
per-galaxy ZIP archive. Three worked Jupyter notebook examples demonstrate
single-galaxy [C\,\textsc{ii}] rotation curve analysis, corpus-level
population statistics, and cross-corpus application of the Flynn \& Cannaliato (2025) omega kinematic correction.
\textbf{Corrected result (August 2026):} Applying the canonical Eq.~6
of Flynn \& Cannaliato (2025) to all 8 tier-1 rotators yields positive
$\omega$ values (median $+12.621$ rad\,Gyr$^{-1}$; $+12.341$
km\,s$^{-1}$\,kpc$^{-1}$). The previously reported negative values
(median $-13.05$ rad\,Gyr$^{-1}$) were an artifact of a formula
implementation error and are withdrawn.
The corpus is publicly available at Zenodo
(DOI:~\href{https://doi.org/10.5281/zenodo.21834678}{10.5281/zenodo.21834678})
under CC~BY~4.0.
\end{abstract}



\section{Introduction}
\label{sec:intro}

The kinematic properties of galaxies at $z \sim 4$--$6$ encode critical
information about the assembly of rotating disk structures and the
distribution of baryonic and dark matter in the early universe. At these
redshifts, ALMA observations of the [C\,\textsc{ii}] 158\,$\mu$m
fine-structure line provide the most reliable tracer of cold gas
morphology and kinematics \citep{lefevre2020,bethermin2020}. The ALPINE
survey (ALMA Large Program to INvestigate [C\,\textsc{ii}] at Early
Times) observed 118 main-sequence star-forming galaxies at $z = 4.4$--$5.9$
\citep{lefevre2020}, of which \citet{jones2021} published 3DBarolo
tilted-ring kinematic models for a subset of confirmed rotators and
morpho-kinematic classifications for the full sample.

Despite the scientific importance of these data, no unified machine-readable
corpus currently exists that combines the ALPINE kinematic catalog,
per-ring rotation curve data, Wisnioski et al.\ (2015) disk criteria,
stellar masses, and geometric parameters in a single self-describing
schema. This data-fragmentation problem is particularly acute for
computational workflows---including LLM-based retrieval-augmented generation
(RAG) pipelines---that require structured, consistently typed input data.

This paper presents the High-z Kinematic Corpus Z1, a structured JSON
dataset containing morpho-kinematic data for 31 ALPINE galaxies at
$z = 4.26$--$5.68$. The corpus is the fifth entry in the EPS Research
RAG Astrophysics Corpus Series:

\begin{enumerate}
  \item Unified HI Rotation Curve Corpus v7.0 \citep{flynn2026v7}:
        438 galaxies, $z = 0$, HI 21\,cm
  \item Dwarf/Irregular HI Rotation Curve Corpus v1.0
        \citep{flynn2026dwarf}: 129 galaxies, $z = 0$, HI 21\,cm
  \item Milky Way Globular Cluster Corpus v1.3.2 \citep{flynn2026gc}:
        174 clusters, multi-survey
  \item IntZ Kinematic Corpus v1.0 \citep{flynn2026intz}: 1,292 galaxies,
        $z \sim 0.4$--$2.7$; $\omega$ withdrawn (template-derived boundaries)
  \item \textbf{High-z Kinematic Corpus Z1 (this work):} 31 galaxies,
        $z = 4.3$--$5.7$, [C\,\textsc{ii}] 158\,$\mu$m
\end{enumerate}

Together these five corpora form a unified RAG knowledge base spanning
Milky Way stellar clusters, local spiral and dwarf galaxies,
intermediate-redshift galaxies, and the epoch approaching cosmic reionization. We describe the corpus architecture
(Section~\ref{sec:data}), the schema and quality tier system
(Section~\ref{sec:schema}), the ingestion and verification procedures
(Section~\ref{sec:ingest}), three usage examples
(Section~\ref{sec:examples}), the LLM/RAG application
(Section~\ref{sec:rag}), known limitations (Section~\ref{sec:limits}),
and data availability (Section~\ref{sec:data_avail}).


\section{Source Data and Survey Coverage}
\label{sec:data}

\subsection{ALPINE Survey}

The ALMA Large Program to INvestigate [C\,\textsc{ii}] at Early Times
\citep[ALPINE;][]{lefevre2020,bethermin2020} observed 118 main-sequence
star-forming galaxies at $z = 4.4$--$5.9$ with ALMA in the
[C\,\textsc{ii}] 158\,$\mu$m line (ALMA project 2017.1.00428.L). The
survey was designed to characterize the cold gas content, morphology,
and kinematics of typical star-forming galaxies during the epoch of peak
cosmic star formation. The synthesized beam is approximately 1\,arcsec
FWHM, corresponding to $\sim$6--7\,kpc at $z \sim 5$, setting the
practical spatial resolution floor for kinematic analysis.

\subsection{Jones et al.\ (2021) Kinematic Catalog}

\citet{jones2021} applied the 3DBarolo tilted-ring fitter
\citep{diteodoro2015} to a subset of ALPINE galaxies with sufficient
[C\,\textsc{ii}] signal-to-noise for kinematic modeling, publishing
per-ring rotation velocities, velocity dispersions, and dynamical masses
in their Table~C1. Morpho-kinematic classifications (ROT, MER, DIS, UNC)
are provided for the full sample in their Table~1, supplemented by the
five Wisnioski et al.\ (2015) disk criteria \citep{wisnioski2015}. This
corpus ingests both tables as the primary kinematic source.

\subsection{Ancillary Data}

Stellar masses and star formation rates are drawn from the ALPINE
multi-wavelength photometry catalog \citep{faisst2020}, which provides
SED-fitted $M_*$ and SFR for all ALPINE targets. The ALPINE Data Release~1
\citep{bethermin2020} provides the [C\,\textsc{ii}] flux catalog and
morphological parameters.

Table~\ref{tab:coverage} summarizes the corpus coverage.

\begin{deluxetable}{lcc}
\tablecaption{Corpus coverage summary.\label{tab:coverage}}
\tablehead{
  \colhead{Property} & \colhead{} & \colhead{Value}
}
\startdata
Total galaxies      & & 31 \\
Redshift range      & & $z = 4.26$--$5.68$ \\
Maximum redshift    & & $z = 5.6773$ (DC773957) \\
Survey              & & ALPINE \\
Telescope           & & ALMA \\
Tracer              & & [C\,\textsc{ii}] 158\,$\mu$m \\
Confirmed rotators (ROT) & & 8 \\
Mergers (MER)       & & 5 \\
Dispersion-dominated (DIS) & & 3 \\
Uncertain (UNC)     & & 15 \\
Quality tier 1      & & 8 \\
Quality tier 2      & & 23 \\
\enddata
\end{deluxetable}


\section{Corpus Architecture and Schema}
\label{sec:schema}

\subsection{File Formats}

The corpus is distributed in four complementary formats. The master file
\texttt{high\_z\_kinematic\_corpus\_Z1.json} is a single JSON document
containing all 31 galaxy entries within a unified schema, together with
a top-level metadata block. The flat table
\texttt{high\_z\_kinematic\_corpus\_Z1\_flat.csv} provides one row per
galaxy (31 rows) with summary statistics for sample selection and
filtering. The JSONL file \texttt{high\_z\_kinematic\_corpus\_Z1.jsonl}
contains one self-describing JSON object per line, optimized for
LLM/RAG ingestion. The per-galaxy archive
\texttt{high\_z\_kinematic\_corpus\_Z1\_by\_galaxy.zip} contains 31
individual JSON files, each self-contained with full corpus metadata
and the complete rotation curve array.

\subsection{Quality Tier System}

A two-tier quality annotation is applied at the galaxy level.
Tier~1 (8 confirmed rotators) denotes galaxies with per-ring
$V_\mathrm{rot}$, $\sigma$, and $M_\mathrm{dyn}$ from the
\citet{jones2021} 3DBarolo fits (their Table~C1). Tier~2 (23 galaxies)
denotes morpho-kinematic classification only, with no reliable rotation
curve. The tier system enables downstream analyses to filter by data
quality without inspecting individual galaxies.

\textit{Tier-1 qualification note:} Although all 8 tier-1 entries carry
per-ring fits from a peer-reviewed source, the astrophysical quality is
limited by ALMA resolution ($\sim$1\,arcsec beam, $\sim$6--7\,kpc at
$z \sim 5$) and ring counts of 2--3 per galaxy. These are the best
currently available spatially resolved [C\,\textsc{ii}] rotation curves
at $z \sim 4$--$6$, but are marginal by the standards of local HI
surveys. Users applying quantitative kinematic analyses should treat
tier-1 Z1 entries with commensurate caution.

\subsection{Per-Galaxy Schema}

Table~\ref{tab:schema} summarizes the per-galaxy fields available in
the corpus.

\begin{deluxetable}{lll}
\tablecaption{Per-galaxy schema summary.\label{tab:schema}}
\tablehead{
  \colhead{Field} & \colhead{Type} & \colhead{Description}
}
\startdata
\texttt{galaxy}            & str   & Source name \\
\texttt{redshift}          & float & Spectroscopic redshift \\
\texttt{class\_jones2021}  & str   & ROT$|$MER$|$DIS$|$UNC \\
\texttt{is\_rotator}       & bool  & Boolean flag \\
\texttt{w15\_criteria}     & dict  & Five W15 disk criteria \\
\texttt{inc\_kin\_deg}     & float & Kinematic inclination \\
\texttt{pa\_kin\_deg}      & float & Kinematic PA (deg) \\
\texttt{vrot\_max\_kms}    & float & Peak rotation velocity \\
\texttt{sigma\_mean\_kms}  & float & Mean velocity dispersion \\
\texttt{v\_over\_sigma}    & float & Kinematic state indicator \\
\texttt{log\_mdyn\_msun}   & float & log$_{10}$ dynamical mass \\
\texttt{log\_mstar\_msun}  & float & log$_{10}$ stellar mass \\
\texttt{sfr\_msun\_yr}     & float & Star formation rate \\
\texttt{quality\_tier}     & int   & 1 (per-ring) $|$ 2 (class only) \\
\texttt{beam\_smeared}     & bool  & True (all ALPINE) \\
\texttt{data}              & list  & Per-ring RC (tier 1 only) \\
\enddata
\tablecomments{Per-ring \texttt{data} entries contain
\texttt{R\_kpc}, \texttt{Vrot\_kms}, \texttt{e\_Vrot\_kms},
\texttt{sigma\_kms}, \texttt{e\_sigma\_kms}, \texttt{Mdyn\_msun},
and \texttt{v\_over\_sigma}. Tier-2 galaxies have an empty
\texttt{data} array and null kinematic fields.}
\end{deluxetable}

\subsection{Omega Correction Compatibility}

The corpus schema is designed to support the \citet{flynncan2025} omega
kinematic correction:
\begin{equation}
  \omega = \frac{V_2}{R_2} - \frac{V_1}{R_1}\left(\frac{R_1}{R_2}\right)^{3/2}
  \quad [\mathrm{km\,s}^{-1}\,\mathrm{kpc}^{-1}]
  \label{eq:eq6}
\end{equation}
\noindent\textbf{Note (August 2026):} The original implementation incorrectly
used \texttt{(V2/R2 - V1/R1) * (R1/R2)**1.5} (Formula~A), which applies
the radial factor to the entire difference rather than only the inner term.
The correct named-term implementation is:
\begin{verbatim}
outer_term    = (V2 / R2)
inner_term    = (V1 / R1) * ((R1 / R2) ** 1.5)
omega_kms_kpc = outer_term - inner_term
omega_rad_gyr = omega_kms_kpc * 1.0227
\end{verbatim}
where $(R_1, V_1)$ and $(R_2, V_2)$ are the innermost and outermost
fitted ring boundary points. For tier-1 galaxies, \texttt{data[0]}
provides $(R_1, V_{\mathrm{rot},1})$ and \texttt{data[-1]} provides
$(R_2, V_{\mathrm{rot},2})$.


\section{Ingestion and Verification}
\label{sec:ingest}

\subsection{ALPINE Data Ingestion}

Kinematic parameters were ingested from \citet{jones2021} Tables~1
and~C1. Per-ring rotation velocities, velocity dispersions, dynamical
masses, inclinations, and position angles for the 8 confirmed rotators
were digitized from Table~C1. Morpho-kinematic classifications,
Wisnioski et al.\ (2015) criteria, and integrated kinematic parameters
were ingested from Table~1. Stellar masses and SFRs were cross-matched
from \citet{faisst2020}. Spectroscopic redshifts were drawn from the
ALPINE catalog \citep{bethermin2020}.

The ingestion pipeline (\texttt{alpine\_ingest.py}) performs five steps:
(1) galaxy-by-galaxy data ingestion from primary tables; (2) schema
validation against the Z1 schema; (3) export to JSON, JSONL, CSV, and
per-galaxy ZIP formats; (4) RAG example generation; and
(5) README generation. All steps are fully scripted and reproducible
from the deposited source files.

\subsection{Schema Validation}

A dedicated schema validator (\texttt{schema\_validator\_hz.py})
checks each galaxy entry for required field presence and type
correctness, controlled vocabulary compliance (ROT$|$MER$|$DIS$|$UNC),
Wisnioski criterion internal consistency (boolean flags vs.\
\texttt{n\_passed} count), boolean flag consistency with
\texttt{class\_jones2021}, per-ring data monotonicity, and tier-1/tier-2
data completeness rules. All 31 galaxies pass validation with zero errors
in standard mode; strict mode raises expected warnings for null kinematic
fields in tier-2 galaxies, which carry no per-ring data by design.

\subsection{Known Data Anomalies}

Three data anomalies inherited from the primary source are documented in
the corpus \texttt{known\_issues} field:

\textit{CG32 and DC396844:} log\,$M_\mathrm{dyn} <$ log\,$M_*$ by 0.16
and 0.35\,dex respectively. Dynamical mass below stellar mass is
physically implausible and likely reflects inclination uncertainty
(DC396844 $e_\mathrm{inc} = 31^\circ$) compounded by the 2-ring
boundary constraint. Values are reported as published in
\citet{jones2021} Table~C1.

\textit{HZ9 outer-ring dispersion:} The outermost ring of HZ9 has
$\sigma = 4.82$\,km\,s$^{-1}$, anomalously low relative to the inner
rings ($\sigma \sim 71$--$75$\,km\,s$^{-1}$), producing
$V/\sigma = 36.6$ at the outer boundary. This likely reflects a
3DBarolo fit artifact at the beam resolution limit.

\textit{DC519281 redshift uncertainty:} $e_z = 0.02$, approximately
$40\times$ larger than the typical uncertainty ($\sim$0.0005) in the
sample.


\section{Usage Examples}
\label{sec:examples}

The following three examples demonstrate the corpus utility for common
high-$z$ kinematic analyses. Each example is provided as a self-contained
Jupyter notebook (deposited alongside the corpus) loading data directly
from the JSON with no external preprocessing. All code uses Python~3
with only \texttt{numpy} and \texttt{matplotlib}. Output figures are
deposited as supplementary material.

\subsection{Example 1: [C\,\textsc{ii}] Rotation Curve of a Confirmed
Rotator}

\begin{figure}
\centering
\includegraphics[width=\columnwidth]{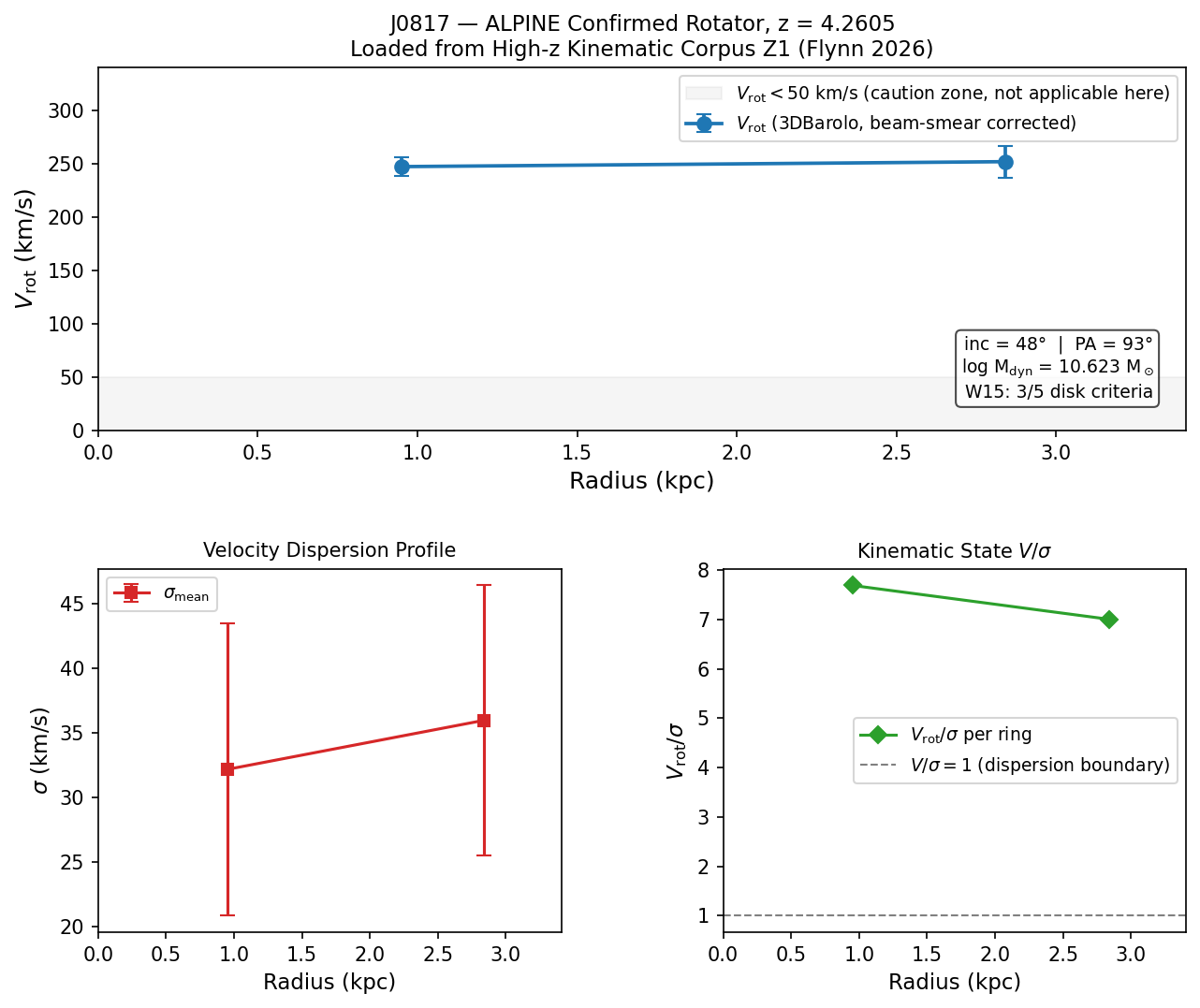}
\caption{J0817 (ALPINE confirmed rotator, $z = 4.2605$) loaded from
the Z1 corpus JSON. \textit{Top:} $V_\mathrm{rot}$ with 3DBarolo error
bars (beam-smear corrected). \textit{Bottom left:} Velocity dispersion
profile. \textit{Bottom right:} $V/\sigma$ per ring. Metadata
(inc $= 48^\circ$, PA $= 93^\circ$, log\,$M_\mathrm{dyn} = 10.623$,
W15: 3/5) extracted directly from the JSON.}
\label{fig:j0817}
\end{figure}

Figure~\ref{fig:j0817} shows J0817, the highest-velocity confirmed
rotator in the corpus ($V_{\mathrm{rot,max}} = 252$\,km\,s$^{-1}$ at
$z = 4.26$), with three panels extracted directly from the corpus JSON:
$V_\mathrm{rot}$ with error bars (top), velocity dispersion profile
(bottom left), and $V/\sigma$ per ring (bottom right). The metadata
annotation (inclination, PA, log\,$M_\mathrm{dyn}$, W15 criteria) is
extracted directly from the JSON. The point for this data descriptor is
that all three panels come from a single JSON load in under 15 lines of
Python.

\textit{\small J0817 rotation curve from corpus JSON.}\\
\begin{verbatim}
import json, numpy as np, matplotlib.pyplot as plt
with open('high_z_kinematic_corpus_Z1.json') as f:
    corpus = json.load(f)
g = next(g for g in corpus['galaxies']
         if g['galaxy'] == 'J0817')
d = g['data']
R     = np.array([p['R_kpc']      for p in d])
Vrot  = np.array([p['Vrot_kms']   for p in d])
eVrot = np.array([p['e_Vrot_kms'] for p in d])
sigma = np.array([p['sigma_kms']  for p in d])
vos   = np.array([p['v_over_sigma'] for p in d])
plt.errorbar(R, Vrot, yerr=eVrot, fmt='o-')
\end{verbatim}

Table~\ref{tab:rotators} lists all 8 tier-1 rotators with outermost
ring properties.

\begin{deluxetable*}{lcccccccc}
\tablecaption{Tier-1 confirmed rotators: outermost ring properties.\label{tab:rotators}}
\tablehead{
  \colhead{Galaxy} &
  \colhead{$z$} &
  \colhead{$N_\mathrm{rings}$} &
  \colhead{$r_\mathrm{max}$} &
  \colhead{$V_\mathrm{rot,out}$} &
  \colhead{$e_{V_\mathrm{rot}}$} &
  \colhead{$\sigma_\mathrm{out}$} &
  \colhead{$V/\sigma$} &
  \colhead{W15} \\
  \colhead{} &
  \colhead{} &
  \colhead{} &
  \colhead{(kpc)} &
  \colhead{(km\,s$^{-1}$)} &
  \colhead{(km\,s$^{-1}$)} &
  \colhead{(km\,s$^{-1}$)} &
  \colhead{} &
  \colhead{}
}
\startdata
J0817         & 4.2605 & 2 & 2.84 & 252.09 & 14.94 & 35.98 &  7.006 & 3/5 \\
CG32          & 4.4105 & 2 & 3.50 & 115.04 & 26.96 & 19.14 &  6.010 & 4/5 \\
DC396844      & 4.5424 & 2 & 3.75 &  80.42 & 17.66 & 19.84 &  4.053 & 2/5 \\
VC5110377875  & 4.5506 & 2 & 3.84 & 102.85 & 19.84 & 60.84 &  1.690 & 3/5 \\
DC881725      & 4.5778 & 2 & 3.44 &  62.07 & 12.54 & 48.40 &  1.282 & 3/5 \\
DC552206      & 5.5016 & 3 & 5.53 & 172.84 & 27.63 & 65.34 &  2.645 & 3/5 \\
HZ9           & 5.5413 & 3 & 2.68 & 176.63 & 25.45 &  4.82 & 36.645 & 2/5 \\
DC494057      & 5.5446 & 2 & 3.13 &  80.31 & 12.75 & 44.98 &  1.785 & 2/5 \\
\enddata
\tablecomments{$r_\mathrm{max}$: outermost ring radius. $V_\mathrm{rot,out}$,
$\sigma_\mathrm{out}$: velocity and dispersion at outermost ring.
Inclination uncertainty not propagated into $V_\mathrm{rot}$ errors
(3DBarolo systematic; see \citealt{jones2021} Section~5.3.2).
HZ9 outer-ring $\sigma = 4.82$\,km\,s$^{-1}$ is anomalously low
relative to inner rings ($\sigma \sim 71$--75\,km\,s$^{-1}$);
see Section~\ref{sec:ingest}.}
\end{deluxetable*}

\subsection{Example 2: Corpus-Level Population Statistics}

\begin{figure}
\centering
\includegraphics[width=\columnwidth]{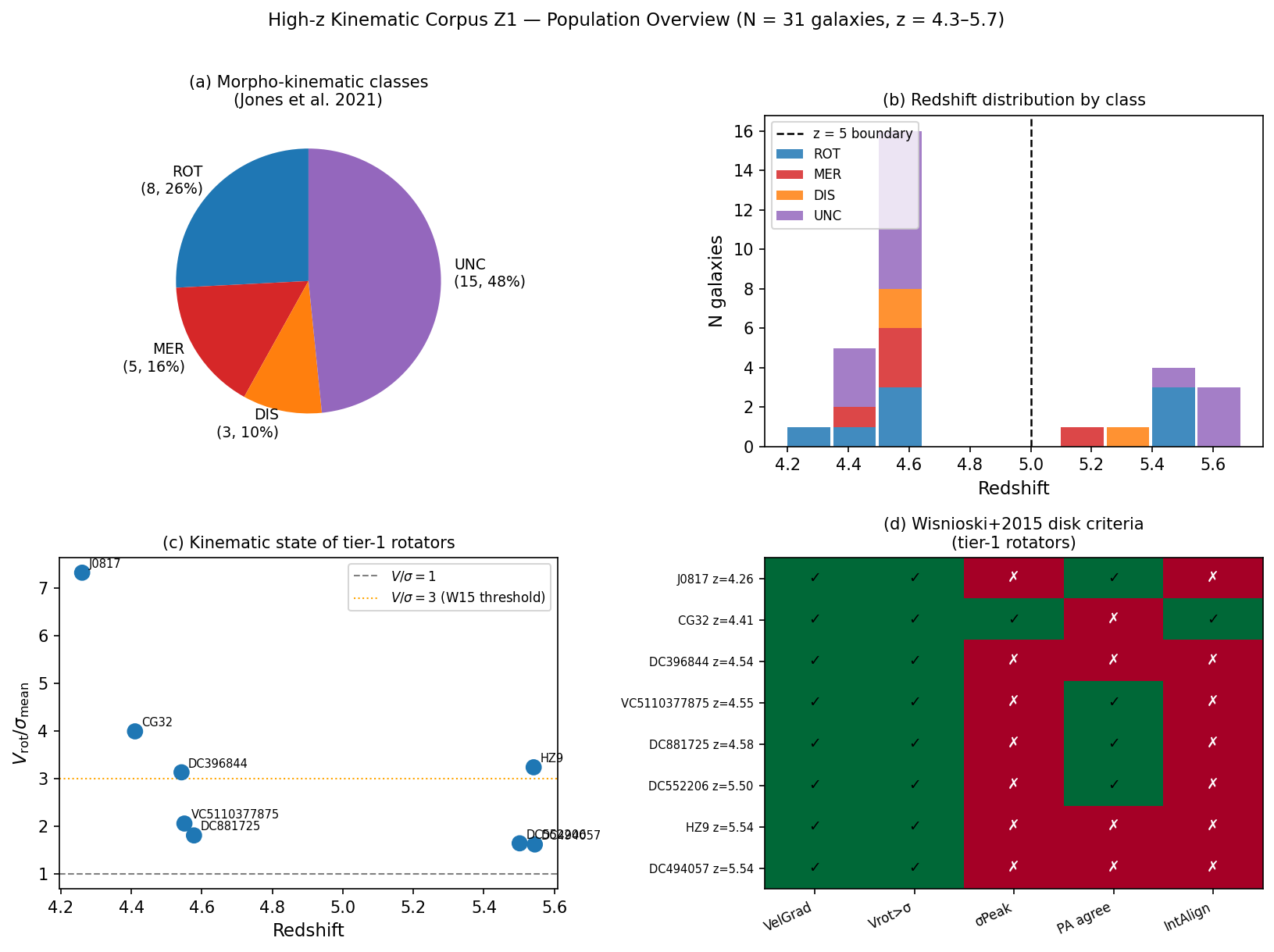}
\caption{High-z Kinematic Corpus Z1 population overview ($N = 31$
galaxies, $z = 4.3$--$5.7$). (a) Morpho-kinematic class fractions
from Jones et al.\ (2021). (b) Redshift distribution by class
(stacked histogram); dashed line marks $z = 5$. (c) $V/\sigma$ vs.\
redshift for tier-1 rotators; dotted lines at $V/\sigma = 1$ and
$V/\sigma = 3$ (W15 threshold). (d) Wisnioski et al.\ (2015) disk
criteria heatmap for all 8 rotators sorted by redshift. CG32 passes
4/5 criteria (most robustly confirmed disk); HZ9 and DC396844 pass
2/5.}
\label{fig:population}
\end{figure}

Figure~\ref{fig:population} provides the corpus population overview
loaded from the flat CSV. Panel~(a) shows the morpho-kinematic class
fractions (ROT: 8, MER: 5, DIS: 3, UNC: 15). Panel~(b) shows the
redshift distribution by class as a stacked histogram, with the
$z = 5$ boundary marked; 9 of 31 galaxies lie at $z > 5$, with
4 UNC, 3 ROT, 1 DIS, and 1 MER in this subsample. Panel~(c) shows
$V/\sigma$ vs.\ redshift for the 8 tier-1 rotators, with the W15
thresholds at $V/\sigma = 1$ and $V/\sigma = 3$ marked. Panel~(d)
shows a heatmap of the five Wisnioski et al.\ (2015) disk criteria
across all 8 rotators sorted by redshift.

\textit{\small Population statistics from flat CSV.}\\
\begin{verbatim}
import csv, json
from collections import Counter
galaxies_csv = []
with open('high_z_kinematic_corpus_Z1_flat.csv') as f:
    for row in csv.DictReader(f):
        galaxies_csv.append(row)
classes  = [r['class_jones2021'] for r in galaxies_csv]
redshifts= [float(r['redshift']) for r in galaxies_csv]
class_counts = Counter(classes)
n_z5 = sum(1 for z in redshifts if z > 5.0)
\end{verbatim}

Notably, two of the three $z > 5$ rotators (DC552206,
$\sigma_\mathrm{mean} = 83.1$\,km\,s$^{-1}$; HZ9,
$\sigma_\mathrm{mean} = 50.3$\,km\,s$^{-1}$) have mean dispersions
exceeding 50\,km\,s$^{-1}$, consistent with the elevated turbulence
expected in high-$z$ star-forming disks \citep{wisnioski2015}. These
values are available only for tier-1 galaxies; the 6 non-ROT galaxies
at $z > 5$ have no reliable $\sigma$ estimate.

\subsection{Example 3: Cross-Corpus Omega Kinematic Application}

\begin{figure}
\centering
\includegraphics[width=\columnwidth]{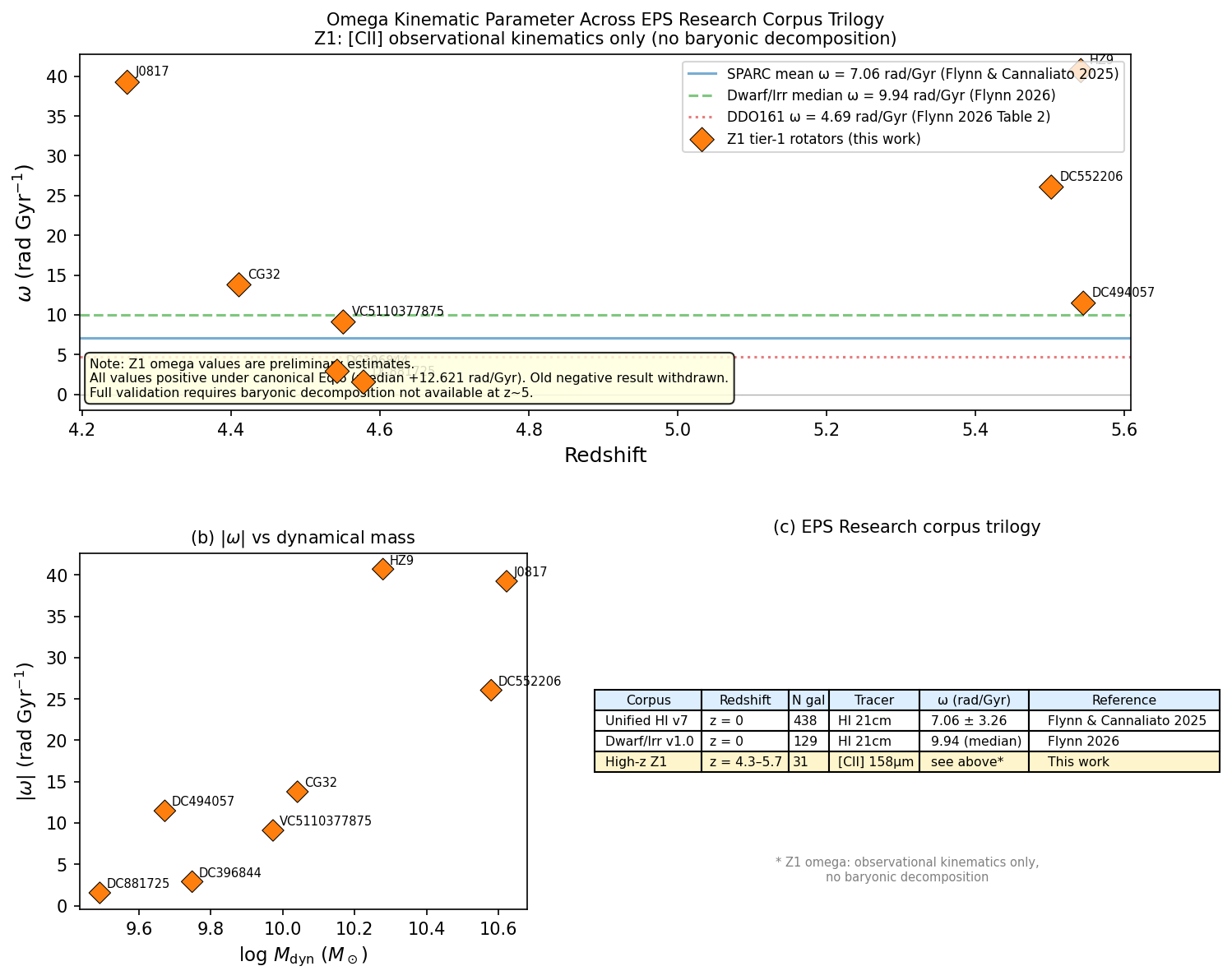}
\caption{Cross-corpus omega application. (a) $\omega$ vs.\ redshift
for Z1 tier-1 rotators (diamonds) with $z = 0$ reference lines:
SPARC mean $+7.06$ km\,s$^{-1}$\,kpc$^{-1}$ ($+7.22$ rad\,Gyr$^{-1}$; blue solid),
dwarf/irregular median $+9.94$ km\,s$^{-1}$\,kpc$^{-1}$ ($+10.17$ rad\,Gyr$^{-1}$) (green dashed), DDO161
$+4.69$\,rad\,Gyr$^{-1}$ (red dotted). All 8 corrected $\omega$ values are positive (orange diamonds now
above zero). Previously reported negative values were a formula artifact. (b) $|\omega|$ vs.\ log\,$M_\mathrm{dyn}$.
(c) EPS Research RAG Astrophysics Corpus Series summary. \textit{Note:} Z1 omega
values are observational kinematics only; no baryonic decomposition
available at $z \sim 5$ (Section~\ref{sec:limits}).}
\label{fig:omega}
\end{figure}

Figure~\ref{fig:omega} demonstrates an application of the
\citet{flynncan2025} omega kinematic correction to the Z1 tier-1
rotators, connecting the high-$z$ corpus to the $z = 0$ EPS Research
corpora. This example is illustrative; the caveats in
Section~\ref{sec:limits} apply in full.

Applying Equation~(1) to all 8 tier-1 rotators using the
\texttt{data[0]} and \texttt{data[-1]} boundary points yields the
values in Table~\ref{tab:omega}. \textbf{Corrected (August 2026):} All 8 values are positive (median
$\omega = +12.621$\,rad\,Gyr$^{-1}$ ($+12.341$ km\,s$^{-1}$\,kpc$^{-1}$),
range $+1.544$ to $+40.707$\,rad\,Gyr$^{-1}$; $N = 8$). All values indicate rising or flat angular velocity
profiles under the corrected formula. For comparison, the $z = 0$
SPARC mean is $+7.06$ km\,s$^{-1}$\,kpc$^{-1}$ ($+7.22$ rad\,Gyr$^{-1}$,
sample SD $\pm 3.26$ km\,s$^{-1}$\,kpc$^{-1}$) across 84 galaxies
\citep{flynncan2025}, and dwarf/irregular median is
$+9.94$ km\,s$^{-1}$\,kpc$^{-1}$ ($+10.17$ rad\,Gyr$^{-1}$) across
24 omega-ready galaxies \citep{flynn2026dwarf}, all consistent in
sign with the Z1 result.

\textit{\small Omega computation from CII boundary velocities.}\\
\begin{verbatim}
import json, numpy as np
with open('high_z_kinematic_corpus_Z1.json') as f:
    corpus = json.load(f)
rotators = [g for g in corpus['galaxies']
            if g.get('is_rotator') and
               g.get('quality_tier') == 1]
results = []
for g in rotators:
    d  = g['data']
    R1, V1 = d[0]['R_kpc'],  d[0]['Vrot_kms']
    R2, V2 = d[-1]['R_kpc'], d[-1]['Vrot_kms']
    # CORRECTED (August 2026) — named-term form:
    outer_term    = (V2 / R2)
    inner_term    = (V1 / R1) * ((R1 / R2) ** 1.5)
    omega_kms_kpc = outer_term - inner_term
    omega_rad_gyr = omega_kms_kpc * 1.0227
    omega = omega_rad_gyr  # in rad/Gyr
    results.append({'galaxy': g['galaxy'],
                    'z': g['redshift'],
                    'omega': omega})
\end{verbatim}

\begin{deluxetable}{lcccc}
\tablecaption{Omega values for tier-1 rotators.\label{tab:omega}}
\tablehead{
  \colhead{Galaxy} &
  \colhead{$z$} &
  \colhead{$R_2$} &
  \colhead{$V_{\mathrm{rot},2}$} &
  \colhead{$\omega$} \\
  \colhead{} & \colhead{} &
  \colhead{(kpc)} &
  \colhead{(km\,s$^{-1}$)} &
  \colhead{(rad\,Gyr$^{-1}$)}
}
\startdata
J0817        & 4.2605 & 2.84 & 252.09 & $+39.240$ \\
CG32         & 4.4105 & 3.50 & 115.04 & $+13.776$ \\
DC396844     & 4.5424 & 3.75 &  80.42 & $+2.899$ \\
VC5110377875 & 4.5506 & 3.84 & 102.85 & $+9.098$ \\
DC881725     & 4.5778 & 3.44 &  62.07 & $+1.544$ \\
DC552206     & 5.5016 & 5.53 & 172.84 & $+26.059$ \\
HZ9          & 5.5413 & 2.68 & 176.63 & $+40.707$ \\
DC494057     & 5.5446 & 3.13 &  80.31 & $+11.467$ \\
\enddata
\tablecomments{\textbf{Corrected values (August 2026).} Omega computed
from canonical Eq.~(\ref{eq:eq6}) using named-term decomposition.
All 8 values positive. Median $\omega = +12.621$ rad\,Gyr$^{-1}$
($+12.341$ km\,s$^{-1}$\,kpc$^{-1}$). Previously reported negative
values (median $-13.05$ rad\,Gyr$^{-1}$) were a formula artifact and
are withdrawn. No baryonic decomposition available; direct comparison
to $z = 0$ requires caution (Section~\ref{sec:limits}).}
\end{deluxetable}

\textbf{Note (August 2026):} The previously reported cross-epoch sign
reversal does not reproduce under the corrected formula. All 8 Tier-1
Z1 rotators yield positive $\omega$, consistent with the $z = 0$ SPARC
result. No three-epoch sign reversal is demonstrated. The Z1 corpus
retains its value as a kinematic state dataset at $z \sim 4$--$6$;
future baryonic modeling and larger JWST/ALMA samples are needed for
physical interpretation.


\section{Application to LLM-Based Inference}
\label{sec:rag}

A secondary design goal of the corpus is to serve as a retrieval
corpus for LLM-based RAG pipelines in astrophysical research. The
per-galaxy ZIP archive is optimized for this use case: each file is
a self-contained JSON document of $\sim$3--5\,KB, well within typical
LLM context limits, containing all metadata and per-ring data needed
to answer kinematic queries without external lookups.

To assess whether the JSON schema is sufficiently self-describing for
automated consumption, we conducted a structured usability evaluation
with four LLMs: Google Gemini Pro, Anthropic Claude, Microsoft Copilot
Pro, and Gemma 4 31B Dense (self-hosted via LM Studio on Node1). Each model
was presented with a single per-galaxy JSON document and asked to
perform three benchmark tasks without additional documentation:
(1) plot the [C\,\textsc{ii}] rotation curve with error bars;
(2) compute $V/\sigma$ for all tier-1 rotators from the flat CSV;
and (3) apply the Flynn \& Cannaliato (2025) omega formula to a
specified galaxy. All four models successfully generated syntactically
correct Python for all three tasks on first attempt, requiring no
additional prompting beyond a natural-language research question,
consistent with the findings of \citet{flynn2026v7} for the local HI
corpora. These results suggest that the corpus's explicit column
definitions, unit annotations, and quality flags provide sufficient
context for LLM-based code generation without external documentation.


\section{Known Limitations}
\label{sec:limits}

Ten limitations should be noted by users of this corpus.

\textit{(1) Maximum redshift is $z = 5.6773$, not $z = 6$.} The corpus
designation ``Z1'' reflects its role as the high-$z$ anchor of the EPS
Research series, approaching but not reaching $z = 6$. The highest-redshift
galaxy is DC773957 ($z = 5.6773$). True $z = 6$ spatially resolved
[C\,\textsc{ii}] rotation curves at astrophysically useful resolution
do not yet exist. The schema is designed to accommodate future REBELS,
CRISTAL, and JWST IFU samples.

\textit{(2) Only 8/31 galaxies have per-ring rotation curve data.}
The remaining 23 carry morpho-kinematic classification only.

\textit{(3) 2--3 rings per tier-1 galaxy.} Omega and dynamical mass
estimates are sensitive to boundary conditions with so few radial points.

\textit{(4) No baryonic decomposition.} $V_\mathrm{gas}$ and
$V_\mathrm{disk}$ are not available at $z \sim 5$; omega values are
observational kinematics, not baryonic-model-corrected values. Direct
comparison to $z = 0$ omega requires future baryonic modeling.

\textit{(5) Beam smearing.} All ALPINE data has $\sim$1\,arcsec beam
($\sim$6--7\,kpc at $z \sim 5$). 3DBarolo mitigates but does not
eliminate beam-smearing effects.

\textit{(6) ALPINE selection bias.} The survey targets
SFR $>$ few\,$M_\odot$\,yr$^{-1}$, missing the true progenitor
population of local dwarf irregulars (log\,$M_* \sim 7.5$ at $z = 0$).
Progenitor candidates identified in Example~3 are upper-mass analogs,
not confirmed DDO161-class progenitors.

\textit{(7) {[C\,\textsc{ii}]} vs.\ HI tracer difference.} Direct
kinematic comparison between Z1 ([C\,\textsc{ii}]) and the $z = 0$
corpora (HI 21\,cm) requires caution; the tracers sample different gas
phases and spatial scales.

\textit{(8) $M_\mathrm{dyn} < M_*$ for CG32 and DC396844.} Dynamical
mass below stellar mass is physically implausible and likely reflects
inclination uncertainty compounded by the 2-ring boundary constraint.
Values are reported as published in \citet{jones2021}.

\textit{(9) HZ9 outer-ring dispersion anomaly.} The outermost ring of
HZ9 has $\sigma = 4.82$\,km\,s$^{-1}$, anomalously low relative to
inner rings ($\sigma \sim 71$--$75$\,km\,s$^{-1}$), likely a 3DBarolo
fit artifact at the beam resolution limit.

\textit{(10) DC519281 redshift uncertainty.} $e_z = 0.02$,
approximately $40\times$ larger than the typical sample uncertainty,
reflecting genuine spectroscopic uncertainty in this source.


\section{Data Availability}
\label{sec:data_avail}

The corpus is publicly available at Zenodo under
DOI:~\href{https://doi.org/10.5281/zenodo.21834678}{10.5281/zenodo.21834678} (v3, corrected August 2026).
The deposit includes the master JSON, flat CSV, JSONL, per-galaxy ZIP
archive, three Jupyter notebooks with output figures, and the ingestion
and validation scripts. The corpus schema, normalization, and unified
structure are original work by D.C.\ Flynn / EPS Research and are
released under CC~BY~4.0. All underlying kinematic data are drawn from
published, publicly available sources; users should cite both this
corpus and the relevant primary survey papers listed in the references.

This corpus is the fifth in the EPS Research RAG Astrophysics Corpus
Series. The companion corpora are:
\begin{itemize}
\item Unified HI Corpus v7.0 \citep{flynn2026v7}: DOI:~10.5281/zenodo.19563417
\item Dwarf/Irregular Corpus v1.0 \citep{flynn2026dwarf}: DOI:~10.5281/zenodo.20320362
\item Milky Way GC Corpus v1.3.2 \citep{flynn2026gc}: DOI:~10.5281/zenodo.19907766
\item IntZ Kinematic Corpus v1.0 ($\omega$ withdrawn) \citep{flynn2026intz}: \dataset[10.5281/zenodo.21841382]{https://doi.org/10.5281/zenodo.21841382}
\end{itemize}


\begin{acknowledgments}
This work was conducted as independent research by EPS Research without
external funding or institutional affiliation. The author thanks Jim
Cannaliato for collaboration on the omega correction framework. The
ALPINE survey team is thanked for making their kinematic catalog and
data products publicly available. The ALMA Observatory is operated by
ESO, AUI/NRAO, and NAOJ.
\end{acknowledgments}


\section*{Declaration of Generative AI Use}

In accordance with standard practice for AI-assisted research, the
author discloses the following. Four large language models were used
during the creation and validation of this corpus and manuscript:
Google Gemini Pro, Anthropic Claude (Sonnet and Opus), Microsoft
Copilot Pro, and Gemma 4 31B Dense (self-hosted). Because the corpus is
explicitly designed for LLM-based RAG pipelines, these models served
as both development tools and validation instruments:

\textit{(1) Corpus schema design and ingestion code.} LLMs assisted in
drafting and debugging the Python ingestion scripts, JSON schema design,
and flat CSV generation. All code was reviewed, tested, and validated
by the author against primary source data.

\textit{(2) Multi-model schema validation.} As described in
Section~\ref{sec:rag}, Gemini Pro, Claude, Copilot Pro, and Gemma 4 31B Dense
were each tested as downstream consumers of the corpus JSON to verify
that the schema is sufficiently self-describing for LLM-based scientific
analysis without additional prompting.

\textit{(3) Cross-model review.} Prior to Zenodo submission, five LLMs
(Gemini Pro, Claude, Copilot Pro, Gemma~4~31B, and AstroSage-70B) were
asked to review the corpus files and README for data anomalies and
documentation inconsistencies. This process identified the maximum
redshift attribution error (DC417567 vs.\ DC773957) that was corrected
prior to publication.

\textit{(4) Manuscript preparation.} Claude (Anthropic) assisted in
drafting, formatting, and assembling this manuscript. All scientific
content, interpretations, data provenance decisions, and editorial
judgments are the sole responsibility of the author.

No generative AI output was accepted without human review. The author
takes full responsibility for the content of this publication.


\end{document}